\documentclass[11pt]{iopart}
\pdfoutput=1
\usepackage{iopams}
\expandafter\let\csname equation*\endcsname\relax
\expandafter\let\csname endequation*\endcsname\relax
\usepackage{amsmath}
\usepackage{graphicx}
\usepackage{wrapfig}

\renewcommand{\mathbf}{\bm}
\newcommand{\ba}{\begin{eqnarray}}
\newcommand{\ea}{\end{eqnarray}}

\newcommand{\omegmo}{\Omega_{{m}}}

\newcommand{\omegko}{\Omega_{{k}}}
\newcommand{\omegmin}{\Omega_{\textnormal{in}}}
\newcommand{\omegmout}{\Omega_{\textnormal{out}}}

\newcommand{\Hperp}{H_{\perp}}
\newcommand{\Hperpo}{H_{\perp_0}}
\newcommand{\Hpar}{H_{\parallel}}

\newcommand{\LCDM}{\Lambda\textnormal{CDM}}

\begin{document}

\title{What's Inside the Cone?}{Numerically reconstructing the metric from observations.}
\author{H.L. Bester$^{1}$, J. Larena$^{2}$, P.J. van der Walt$^{3}$ and N.T. Bishop$^{4}$}
\address{Department of Mathematics\\
Rhodes University\\
Grahamstown 6140\\
South Africa}

\eads{$^{1}$\mailto{g07b1135@campus.ru.ac.za}, $^{2}$\mailto{j.larena@ru.ac.za}, $^{3}$\mailto{p.vanderwalt@ru.ac.za}, $^{4}$\mailto{n.bishop@ru.ac.za}}

\begin{abstract}
We investigate the possibility of using Gaussian process regression to smooth data on the current past null-cone for use as the input to a relativistic integration scheme. The algorithm we present is designed to reconstruct the metric of spacetime within the class of spherically symmetric dust universes, with or without a cosmological constant. Assuming that gravity is well described by General Relativity, we demonstrate how the algorithm can be employed to test the Copernican principle based on currently available observations. It is shown that currently available data is not sufficient for a conclusive result. The intrinsic noise present in realistic data presents a challenge for our smoothing algorithm and we discuss some of its limitations as well as possible extensions to it. We conclude by demonstrating how a direct determination of the cosmological constant is possible using redshift drift data.
\end{abstract}

\maketitle


\section{Introduction}
The large scale geometry of the Universe is usually described by the spatially homogeneous and isotropic \emph{Friedmann-Lema\^itre-Robertson-Walker} (FLRW) metric. This standard approach is very successful, since the concordance model accurately accounts for most if not all current cosmological observations. However, it crucially relies on the \emph{Copernican principle} (CP), which states that we are not located at a preferred place in the universe. This is considered the last metaphysical ingredient in an otherwise physically grounded theory (see \cite{Clarkson:2012bg} for a recent review on the subject). Consistency of the scientific method requires spatial homogeneity to be confirmed with an observational methodology. \\
The rationale behind the observational formalism is this: if the universe really is FLRW, then it should be possible to confirm it by relaxing our symmetry assumptions and relying on astrophysical and statistical modelling techniques instead. This approach, sometimes referred to as the \emph{inverse problem}, is based on the \emph{Observational Cosmology} programme of Ellis and coworkers (initiated in \cite{Ellis1985} with further developments in \cite{Stoeger92.0, Stoeger92.1, Stoeger92.2, Stoeger92.3, Stoeger94, Maartens96, Stoeger97, araujo00, Araujo2009zh, araujo09.1, araujo10, araujo11}) and also follows related work on inhomogeneous cosmologies such as \cite{Bolejko:2011ys,Hellaby:2008pp,McClure:2007hy,Lu:2007gr,Mustapha1998,Clarkson:2011br,Clarkson:2010uz}. Although the spirit and the methods we use are very similar, in this paper we aim to elucidate the importance of using not only the best-fit to observations but the full statistical information available on our current past null-cone (past null-cones of the central observer at an arbitrary time along the central worldline will be abbreviated by PNC throughout). The statistical information is obtained in a Bayesian way by using Gaussian process regression to smooth observations on the current past null-cone (hereafter PNC0). The smoothing method is completely model independent and doesn't presuppose any physics. Before the data can be used as initial data for the numerical integration scheme it has to be screened to identify the physically realistic combinations of data points: we enforce the null energy condition with the help of a general relativistic constraint for this purpose. The data that passes the screening process is then fed into the integration scheme. This effectively solves the \emph{Einstein field equations} (EFE) for spherically symmetric dust universes (or \emph{Lema\^{i}tre-Tolman-Bondi} (LTB) models \cite{lemaitre33,tolman34,bondi47}) in observational coordinates and allows us to reconstruct the causal history of the universe. In particular we are able to evaluate the matter shear along the PNCs and use it to formulate a test for the CP. Let us emphasise that the algorithm we propose here is restricted to LTB models that may include a cosmological constant. Thus it is able to detect signatures of cosmic scale radial inhomogeneities but it does not address the more general averaging or coarse-graining issues (for comprehensive reviews of these topics, see \cite{Clarkson:2011zq,Buchert:2011sx} and references therein).\\
The problem of reconstructing the metric in the interior of the PNC is formulated as a reversed \emph{characteristic initial value problem} (CIVP) and consists of two phases viz. setting up data on the PNC0 and then integrating the EFEs for the interior. The numerical code developed in \cite{prd2010} and \cite{prd2012} was adapted to facilitate the second phase of the problem. This is a predictor-corrector code based on a 3D code developed for gravitational radiation in \cite{prd1997} and \cite{Bishop:1999yg}.  The codes in \cite{prd2010, prd1997,Bishop:1999yg} were directly based on Bondi-Sachs coordinates \cite{Bondi1962}, where the radial coordinate is the angular diameter distance. However, for cosmology on the PNC, the diameter distance is only useful prior to the apparent horizon after which it becomes multi-valued as the PNC refocuses. In [9], an affine parameter was introduced as a radial coordinate specifically to handle calculations beyond the apparent horizon. Special precautions are required to deal with the fact that the null affine parameter is necessarily not comoving (see section \ref{sec:NonComAff} for details). \\
We note that, as input, the code requires three functions on the PNC0: the angular diameter distance, the $u_1$ component of the four velocity of dust and the relativistic energy density of dust. It is extremely difficult to obtain the density directly from observations in a model independent way. The density is therefore reconstructed by relying on a general relativistic constraint that uses derivatives of the distance/redshift and age/redshift relations. The uncertainty inherent in the reconstruction of successive derivatives of these relations leads to large errors in the density. We therefore anticipate that a model independent measure of the density, albeit only for redshifts up to $z \approx 0.4$, will greatly benefit the observational cosmology programme.\\
As far as the initial data is concerned, the challenge is to obtain model independent data i.e. data that does not require the assumption of an FLRW cosmology for its validity. In this paper we use Supernova Ia luminosity data from the Union 2.1 compilation \cite{Suzuki:2011hu} to reconstruct the angular diameter distance. Although not ideal, age data from Cosmic Chronometers \cite{Moresco:2012by} is used to reconstruct the longitudinal Hubble rate \footnote{See section \ref{ExpRate} for a discussion on the model dependence of this data set}. To smooth these data sets we make use of the publicly available \emph{Gaussian processes in Python} (GaPP) package \cite{Seikel:2012uu}.\\
The paper is organised as follows: section~\ref{sec:formalism} outlines the observational formalism in spherically symmetric dust universes and introduces the numerical integration scheme used to solve for the interior of the PNC. In section~\ref{sec:setdata} we indicate what observations are required to fix the initial conditions for the integration scheme and outline the method used to smooth these data sets. The main results of the paper are presented in section~\ref{sec:Res} where we formulate a test of the CP. In section~\ref{sec:discussion} we finish off the paper with a discussion of possible extensions to the algorithm presented here.\\
We use the signature $(-+++)$ of the metric throughout and work with geometrised units where $c = G = 1$. An overdot ($\dot{~}$) and prime ($'$) represents derivatives w.r.t. proper time along our central worldline ($w$) and the affine parameter ($\upsilon$), respectively. A subscript zero could refer either to quantities evaluated on the current time slice or on the PNC0, this should be clear from the context.

\section{Formalism}\label{sec:formalism}

\subsection{Fluid description}\label{sec:EneDes}
In this paper we assume that the matter content of the late time universe can be sufficiently well described by treating it as a single perfect fluid modelled as a spherically symmetric dust distribution with energy momentum tensor
\begin{equation}
T_{ab} = \rho u_a u_b, \qquad u_au^a = -1.
\label{PerEMT}
\end{equation}
Here $\rho$ is the relativistic energy density of dust and $u^a = \frac{dx^a}{d\tau}$ is the normalised fundamental fluid 4-velocity. Following \cite{ellis2012relativistic}, the observational approach makes use of two fundamental assumptions about the constituents of the universe:
\begin{itemize}
\item  baryons and cold dark matter (CDM) are comoving (i.e. $u^a_{B} = u^a_{CDM}$ and therefore, $u^a_{CDM}$ is inferred from $u^a_{B}$ observations),
\item the value of the cosmological constant $\Lambda$ is known independently of cosmological observations.
\end{itemize}
Interpreting $\Lambda$ as vacuum energy arising from statistical fluctuations of some quantum field leads to a value that is at present about a hundred and twenty orders of magnitude larger than what is expected from cosmological considerations (see for example \cite{peter2009primordial}). The implication for observational cosmology therefore, is that we cannot yet accomplish the observational ideal of reconstructing the observer metric purely from model independent observations. Moreover, considering the discrepancy in the cosmological and field theoretical value of $\Lambda$, this will probably remain true for some time to come. An alternative, although not independent of cosmological observations, is to observe $\Lambda$ directly using redshift drift data (see section~\ref{sec:discussion} for an illustration). Redshift drift could in principle be used to constrain many cosmological parameters directly (see for instance \cite{Loeb:1998bu}, \cite{PhysRevD.75.062001}, \cite{2007MNRAS.382.1623B}, \cite{PhysRevLett.100.191303}) but such data will not be available for some time.  

\subsection{Observational coordinates}
\begin{figure}
\centering
\includegraphics[width = 0.8\textwidth]{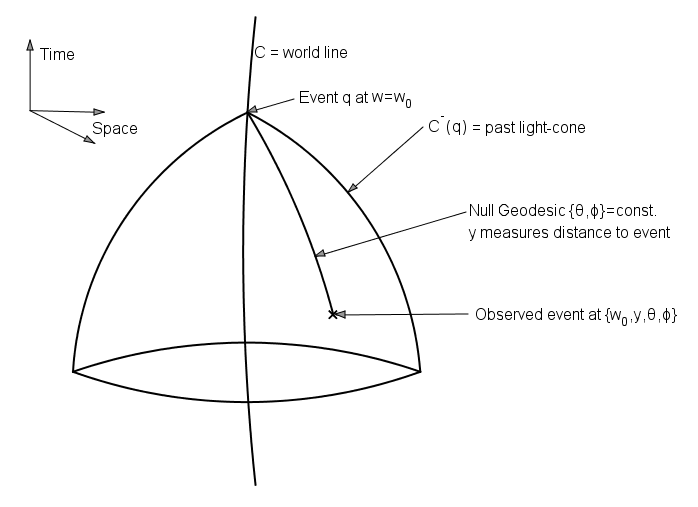}
\caption{Observational Coordinates}
\label{fig:ObsCor}
\end{figure}

We make use of observational coordinates as introduced in \cite{Ellis1985} while mostly following the methodology as presented in \cite{ellis2012relativistic}. Accordingly, let $x^a = (w,y,\theta,\phi)$ be observational coordinates defined as follows (see figure \ref{fig:ObsCor}):
\begin{itemize}
\item Let $w = $const. define the PNCs $C^-(w)$ of the events along the observer worldline $C$ normalised so that $w$ on $C$ is the proper time along the central worldline i.e. $ds^2|_C = -dw^2$. Then the $C^-(w)$ are generated by the past directed geodesic ray 4-vector (note that $k^a$ is not defined at the vertex $q$):
\begin{equation}
 k_a = \partial_a w, \quad k^a = \frac{dx^a}{d\upsilon}, \quad k^ak_a = 0 = k^b\nabla_b k^a,
\label{4vec}
\end{equation}
where $\upsilon$ is an affine parameter. The affine freedom is fixed by choosing $\upsilon = 0$ on $C$ and by the central condition
\begin{equation}
k_au^a|_C = 1.
\label{CenCon}
\end{equation}
\item Let $y$ be a radial coordinate down the null geodesics that rule the $C^-(w)$. $y$ then measures spatial distance from $C$ as well as temporal distance from the observer at the vertex $q$. Examples include: the affine parameter $y = \upsilon$, the redshift $y = z$ or the angular diameter distance $y = D$. In practical applications, observations are generally reported as functions of redshift so the $y(z)$ relation is critical.
\item Let $x^I = (\theta,\phi)$ be angular coordinates based on a parallelly propagated tetrad along $C^-(w)$ (where capital Latin indices run through $2,3$ only). These are chosen so that they label the null geodesics in each PNC i.e. $k^a \partial_a x^I = 0 = k^I$. Clearly we have:
\begin{equation}
k^a = B^{-1}\delta^a_{\ 1}, \quad \mbox{where} \quad B := \frac{d\upsilon}{dy}, \quad \mbox{and} \quad k_a = \delta^{\ 0}_a.
\label{GenGeo}
\end{equation}
\end{itemize}
In a general space-time the observed expansion rate has the covariant definition \cite{1969CMaPh..12..108E}
\begin{equation}
H_{obs} = \frac{1}{3}\Theta - A_ae^a + \sigma_{ab}e^ae^b,
\label{CovExpRate}
\end{equation}
where $\Theta$ is the isotropic expansion scalar, $A_a$ is the acceleration dipole which vanishes in pure dust universes, $\sigma_{ab}$ is the shear quadrupole and $e^a$ is the spatial direction of observation. The relation required to convert observations from functions of $z$ to functions of the affine parameter is found by making use of the geometric optics approximation to decompose the ray 4-vectors into the screen space. It follows that \cite{Clarkson:2010uz}
\begin{equation}
\frac{dz}{d\upsilon} = (-u_ck^c)^2\left(\frac{1}{3}\Theta - A_ae^a + \sigma_{ab}e^ae^b\right),
\label{nuz1}
\end{equation}
where $-u_ak^a$ is related to the energy of the ray which, combined with the central condition \eqref{CenCon}, can be interpreted as $u_ak^a = 1+z$. Thus the required relation follows from \eqref{CovExpRate} and \eqref{nuz1} as
\begin{equation}
\frac{dz}{d\upsilon} = (1+z)^2H_{obs}, \quad \Rightarrow \quad \upsilon(z) = \int_0^z \frac{dz'}{(1+z')^2H_{obs}(z')},
\label{nuz}
\end{equation}
where $\upsilon(0) = 0$ has been used. Moreover, any quantity $X$ evaluated along the ray bundle satisfies \cite{Clarkson:2011br}
\begin{equation}
\frac{dX}{d\upsilon} = (1+z)^2H_{obs}(z) \frac{dX}{dz}.
\label{dnuz}
\end{equation}
By isotropy around the central worldline we also have that $\rho=\rho(w,y)$, $u^{a}=u^{a}(w,y)$ and in addition $u^{\theta}=u^{\phi}=0$. The metric in isotropic observational coordinates then takes the simple form
\begin{equation}
ds^2 = -A(w,y)^2dw^2 + 2B(w,y)d\upsilon dw + D(w,y)^2d\Omega^2.
\label{AffObsCoo}
\end{equation}
Further we choose $y = \upsilon$ so that $B = 1$. This choice is motivated by the fact the $\upsilon(z)$, as opposed to $D(z)$, is always a one to one relation. In particular, with $y = \upsilon$ it is possible to go beyond the apparent horizon without resorting to series expansions. However a subtlety arises since it is not possible to choose $\upsilon$ as a comoving coordinate if $B = 1$ is to be maintained in the interior of the PNC. Since the observable universe was smaller in the past the maximum value of $\upsilon$ decreases on previous PNCs. As explained in section \ref{sec:NonComAff} we take special precautions to identify the causally connected region in the interior of the PNC.

\subsection{Integration scheme}\label{sec:IntSch}
For the integration the metric is rewritten with $A^2(w,\upsilon) = (1+\frac{W(w,\upsilon)}{D(w,\upsilon)})$ \cite{prd2012} (a form  more reconcilable with the original Bondi-Sachs form \cite{Bondi1962} used to study gravitational radiation) i.e.
\begin{equation}
ds^2 = -\left(1+\frac{W(w,\upsilon)}{D(w,\upsilon)}\right) dw^2 + 2dwd\upsilon + D(w,\upsilon)^2d\Omega^2.
\label{CIVPMet}
\end{equation}
Regularity at the vertex then imposes that the function $W$ must tend to zero more rapidly than $D$ as $\upsilon \rightarrow 0$. Substituting the metric \eqref{CIVPMet} into the EFEs, using the  form $R_{ab} = \kappa(T_{ab} - \frac{1}{2}Tg_{ab}) + \Lambda g_{ab}$, leads to
\begin{eqnarray}
\hspace{-1.5cm} D'' &=& -\frac{1}{2}\kappa D \rho (u_1)^2, \label{RNU} \\
\hspace{-1.5cm} \dot{D}' &=& \frac{1}{2D} \left[1-W'D' - DD'' - W D'' - 2\dot{D}D' - (D')^2 - \frac{1}{2}\kappa \rho D^2 - \Lambda D^2\right], \label{RW} \\
\hspace{-1.5cm} W'' &=& \frac{W}{D}D'' - 4\dot{D}' - 2\kappa(u_0u_1\rho - \frac{1}{2}\rho)D - 2\Lambda D,  \label{WNU}\\
\hspace{-1.5cm} \mbox{where}&:& D(0) = W(0) = W'(0) = \dot{D}(0) = 0~\mbox{and}~D'(0)=1.
\end{eqnarray}
These can be considered as constraint equations that need to be solved on each PNC. To evolve the system to the next PNC the values of $\dot{u}_{1}$ and $\dot{\rho}$ need to be specified. These follow from the conservations equations $\nabla_bT^b_{\ a} = 0$:
\begin{eqnarray}
\hspace{-1.5cm} \dot{u}_{1} &=& -\frac{1}{u_1}\left[(A^2 u_1 + u_0)(u_{1})' + AA' (u_1)^2\right], \label{U1W}\\
\hspace{-1.5cm} \dot{\rho} &=& -\frac{1}{u_1}\big[\rho(\dot{u}_{1} + (u_{0})' + 2AA''u_1 + A^2 (u_{1})' + 2\frac{D'}{D}(u_0+A^2 u_1)+2\frac{\dot{D}}{D}u_1) \nonumber \\
\hspace{-1.5cm}  &\ & +\rho'(u_0 + A^2 u_1) \big], \label{RHOW}
\end{eqnarray}
where we show the equations with $A^2 = (1+\frac{W}{D})$ for brevity.\\
These equations form a system that can be solved in the order \eqref{RNU}, \eqref{RW} and \eqref{WNU}. The normalisation of $u^a$ gives $u_0$ in terms of $u_1$ via:
\begin{equation}
u_0 = -\frac{1}{2u_1} -\frac{A^2 u_1}{2}.
\label{IntNorCon}
\end{equation}
Then, using \eqref{U1W} and \eqref{RHOW} evolves the system to the next PNC where the procedure can be repeated until the domain of integration is covered. This scheme was implemented and validated against ideal data in \cite{prd2012} where a more detailed description is presented \footnote{The equations in \cite{prd2012} are written in terms of a future null cone and the integration direction was adapted to have an effective PNC. As a result there are various terms with opposite signs.}. It effectively solves the inverse problem for the class of spherically symmetric cosmological models which can include a cosmological constant. Since it is insensitive to both the sign of the spatial curvature and the apparent refocussing of the angular diameter distance, it applies generally to a large class of models and over a very wide redshift range.  

\subsection{Non-comoving radial coordinate}\label{sec:NonComAff}
Since $\upsilon$ is not a comoving coordinate (otherwise $B \neq 1$ in the interior of the PNC), a subtlety arises in going from one PNC to the next. Although the $\upsilon$ coordinate is causally aligned in the radial direction, the maximum extent of the calculation on older PNCs is not controlled to remain causally aligned with the maximum value on the PNC0. To take account of the causal boundary the domain of calculation has to be reduced. This is done by calculating the path of an incoming characteristic from the maximum value of $\upsilon$ on the PNC0 ($\upsilon_{max}$) into the interior of the PNC. Numerically this is implemented by integrating the interior in two runs. On the first run the grid is written with $\upsilon_{max}$ fixed. Then a null cut-off is computed using:
\begin{eqnarray}
ds^2 &=& 0 = -\left(1+\frac{W}{D}\right)dw^2 + 2dwd\upsilon, \\
dw &\neq& 0 \Rightarrow \upsilon - \upsilon_1 = \frac{1}{2}\int_{w_1}^{w} \left(1 + \frac{W}{D}\right)dw^*.
\end{eqnarray}
The value of $\upsilon_{max}$ can then be updated iteratively by first estimating the next affine parameter maximum $\upsilon_{next}$ as
\begin{equation}
\upsilon_{next} = \upsilon_{max} + \frac{1}{2} \Delta w,
\end{equation}
where $\Delta w$ is negative. Since the $W(\upsilon)$ relation is known on each PNC we can estimate $W(\upsilon_{next})$ and recalculate
\begin{equation}
\upsilon_{next} = \upsilon_{max} + \frac{1}{2}\left(1 + \frac{W(\upsilon_{next})}{D(\upsilon_{next})}\right)\Delta w.
\end{equation}
This is repeated until the value of $\upsilon_{next}$ converges. Since the value of $\upsilon_{next}$ does not necessarily fall on a grid point, the value of $W(\upsilon_{next})$ is calculated by linearly interpolating between the closest grid points and a specific $w_i$ grid line.

\section{Setting the initial data}\label{sec:setdata}
\subsection{Observables}
\begin{figure}
\includegraphics[width=\textwidth]{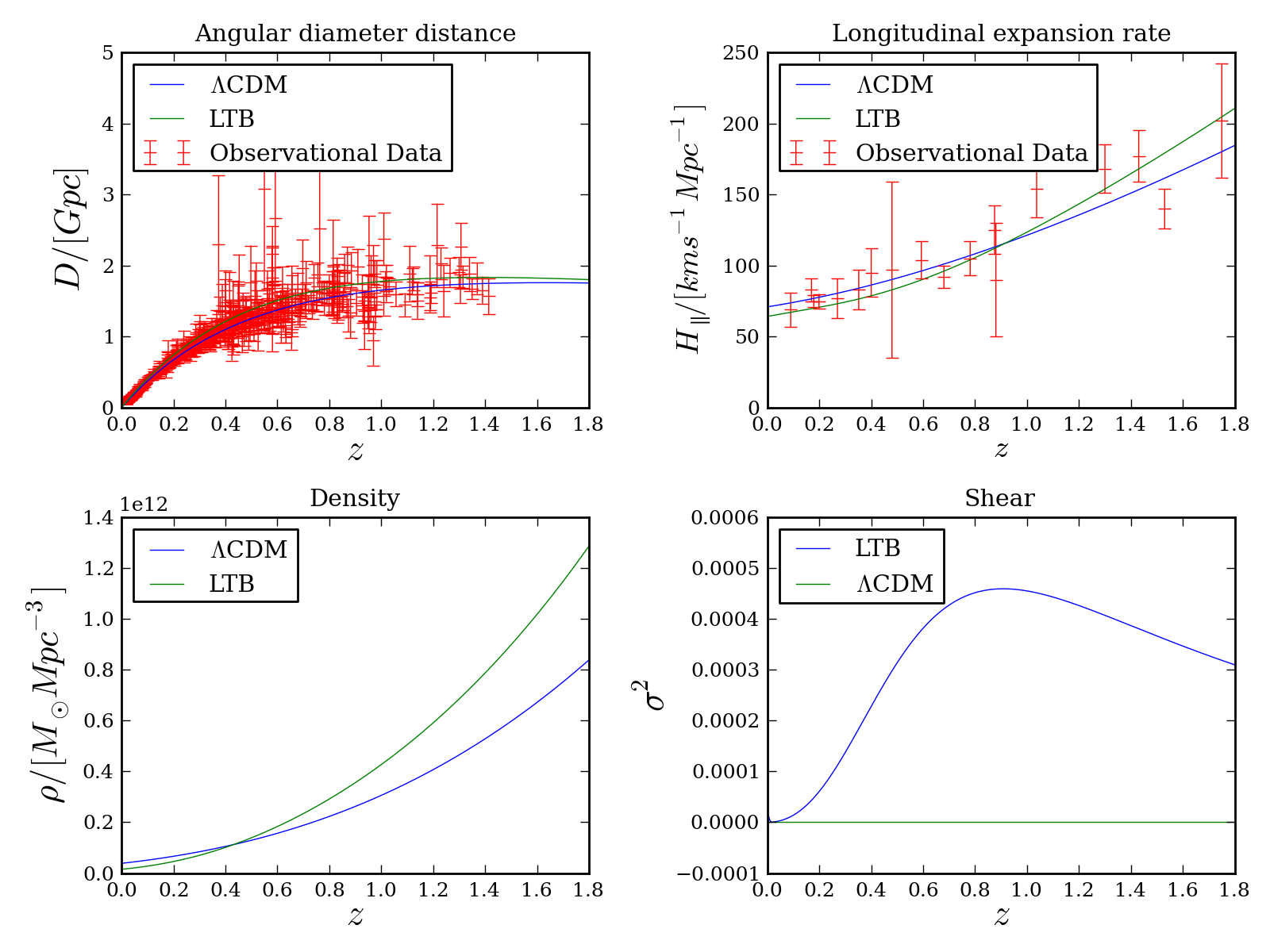}
\caption{The observational data compared to both the LTB and $\Lambda$CDM models.}
\label{fig:ModzRel}
\end{figure}

The functions that need to be specified on the initial PNC are: the angular diameter distance $D(\upsilon)$, the $u_1(\upsilon)$ component of the four velocity, and the density $\rho(\upsilon)$. Since LTB models have two free functional degrees of freedom we need two independent data sets to fix these three functions (see \ref{fig:ModzRel} for a plot of the data sets we use in this paper). In this section we show how all of these quantities can be derived from luminosity and age data (the density follows from the dynamical link provided by the field equations). In addition to these three functions we need to specify two further parameters. One is the value of $\Lambda$\footnote{Treating $\Lambda$ as some dynamical form of dark energy instead of a constant introduces another functional degree of freedom. In this case it is necessary to specify three free functions on the PNC0.}, as discussed in section~\ref{sec:EneDes}, this is currently not feasible. However we show in section~\ref{sec:discussion} how a direct determination of $\Lambda$ is possible using redshift drift data. To set the starting point of the integration we also need to know the current age of the universe along our wordline. Since the aim in this paper is merely to illustrate the feasibility of the algorithm we set this value to $w_0 = 13.6$ Gyr for all the integrations. In the models we consider here the shear is not significantly affected by changing its value slightly (i.e. by $\pm$ 1.5 Gyr).\\
Given the uncertainty in $H_0$ and $\Lambda$, realistic tests of the CP require these parameters to be marginalised over. This issue has been left for future research but see section~\ref{sec:discussion} for a discussion about such a possibility. \\
Further, the fact that $\upsilon$ is the radial coordinate on the grid makes it difficult to incorporate the uncertainty that results from using \eqref{nuz} with the $H_{obs}(z)$ data. We therefore use the most probable value, denoted $\bar{\upsilon}(z)$, as the coordinate on the grid. The uncertainty in $\upsilon(z)$ means that the $\upsilon \leftrightarrow z$ conversion introduces a further uncertainty when interpreting observables as a function of $\upsilon$. Since we only integrate the most probable functions' values on the current PNC this effect (which we anticipate to be significant) is not reflected in our results. Although this might seem like a significant limitation of the algorithm, it can actually be dealt with quite neatly (see section~\ref{sec:discussion}).

\subsubsection{Velocity:~~~}
The fact that $u^ak_a = 1+z = u^0$ and the form of the metric \eqref{AffObsCoo} gives the required component as $u_1 = 1+z$. Thus, ignoring peculiar velocities, its value at each $z$ is given exactly and the only uncertainty in $u_1$ results from the $\upsilon \leftrightarrow z$ conversion. Note however that the integration scheme is very sensitive to errors in $u_1$, which tend to grow rapidly as the integration proceeds.

\subsubsection{Distance:~~~}
We make use of standard candles to constrain the absolute luminosity of a source which is assumed to be emitting isotropically. The luminosity distance $D_L$ is then observable in terms of the distance modulus $\mu$ as
\begin{equation}
\mu = m - M = 5\log_{10}D_L + \mbox{const}.
\end{equation}
Here $m$ is the apparent magnitude and $M$ is the absolute magnitude of the source. We assume that the constant is fixed in the calibration so that the distance modulus vanishes at 10pc i.e.
\begin{equation}
\frac{D_L}{10\mbox{pc}} = 10^{0.2\mu}.
\end{equation}
The angular diameter distance $D$ follows from the distance duality relation \cite{Ellis1971} (assumed to be exact) by
\begin{equation}
D_L = (1+z)^2 D.
\end{equation}
The Union 2.1 data set \cite{Suzuki:2011hu} contains distance modulus vs redshift data from 580 type Ia supernovae (a review of supernovae observations is given in \cite{Astier:2012bq}). Uncertainties in $\mu(z)$ are propagated to $D(z)$ by assuming that errors in $\mu(z)$ are normally distributed.

\subsubsection{Expansion Rate:~~~}\label{ExpRate}
It is convenient to use the 1+3 version of the metric
\begin{equation}
ds^2 = -dt^2 + X^2(t,r)dr^2 + R^2(t,r)d\Omega^2\,,
\label{LTBmetric2}
\end{equation}
where $t$ is the comoving synchronous time and $r$ is the radial coordinate in the spatial hypersurfaces. The radial and angular scale factors are related by
\begin{equation}
X(t,r) = \frac{\partial_r R(t,r)}{\sqrt{1 - k(r)}}.
\label{XvA}
\end{equation}
Note that the curvature $k(r)$ is not constant but is instead a free function. Using these two scale factors we define the perpendicular and longitudinal expansion rates respectively
\begin{equation}
\Hperp= \Hperp(t,r) \equiv \frac{\partial_t X}{X},~~~~\Hpar=\Hpar(t,r) \equiv \frac{\partial_t R}{R}.
\label{Hdefs}
\end{equation}
Using the property $k_a k^a = 0$ for radial null-vectors shows that $k^0 = \pm  X(t,r) k^1$, where the negative corresponds to an incoming ray. Further the null geodesic equation $k^b\nabla_b k^0 = 0$ then reads
\begin{equation}
\frac{dk^0}{d\upsilon} =- (k^0)^2H_{||} \label{geot}.
\end{equation}
Now using $u_ak^a = (1+z) ~ \Rightarrow k^0 = -(1+z)$ we find the $\frac{dz}{d\upsilon}$ relation by differentiating $k^0$
\begin{equation}
\frac{dk^0}{d\upsilon} = -\frac{dz}{d\upsilon} =- (1+z)^2H_{||}.
\label{geo0}
\end{equation}
By comparison with \eqref{nuz} we can identify $H_{||} = H_{obs}$ in LTB. Thus by \eqref{dnuz} we identify
\begin{equation}
k^0 = \frac{dt}{d\upsilon} = (1+z)^2H_{||}(z)\frac{dt}{dz}, \quad \Rightarrow \quad H_{||} = \frac{-1}{1+z}\frac{dz}{dt} \label{Hz}.
\end{equation}
This shows that the longitudinal expansion is given by a simultaneous measure of the redshift and its proper time derivative. The proper time derivative of $z$ can in principle be obtained from a differential dating of old passively evolving galaxies where a model of galactic evolution can be used to infer the change in proper time between galaxies observed at different redshifts. We use the expansion rate data from cosmic chronometers (CC) \cite{Moresco:2012by} \footnote{BAO data is not used since it relies on FLRW perturbation theory}. The validity of using CC with the relation \eqref{Hz} to reconstruct $H_{obs}(z)$ has recently been challenged in \cite{dePutter:2012zx}, where it is pointed out that in an inhomogeneous universe red-envelope galaxies may have a formation time that depends on position. This criticism is valid, thus in order to faithfully use $H_{||}(z)$ data from \eqref{Hz} in an inhomogeneous universe it should be accounted for as systematic errors in the data. In absence of more information on these systematics, we will assume that they are negligible compared to other sources of errors (such as the reconstruction of derivatives for example).

\subsection{Derived quantities}
\subsubsection{Density:~~~}
The density follows from the general relativity constraint
\begin{equation}
k^ak^bR_{ab} = k^ak^b\left(\kappa(T_{ab} - \frac{1}{2}Tg_{ab}) + \Lambda g_{ab}\right),
\end{equation}
which in observational coordinates reads
\begin{equation}
\frac{-2D''}{D} = \kappa \rho (u_1)^2,
\label{density}
\end{equation}
In principle this relation gives an indirect method to compute the density down the PNC, treating $D(z)$ as measurable, and then using $dz/d\upsilon$ from \eqref{geo0} to construct $D(\upsilon)$ and thus $D''$. It should be kept in mind that computing the density this way strongly depends on a number of assumptions. Firstly, the relation is only valid in spherical symmetry or exactly isotropic observations. It also depends on the astrophysical models employed to get the $D(z)$ and $H_{||}(z)$ data and ultimately relies on the validity of general relativity. Because of these implicit assumptions, and the difficulties involved in smoothing the data without imposing a cosmological model (see sections~\ref{sec:GaPP}-\ref{sec:origin}), it may be optimistic to believe that the density obtained in this way is realistic. It is however the only model independent measure of the density available as our current understanding of how the luminosity function relates to number counts relies on the assumption of a cosmological model \cite{Iribarrem:2012yz,Iribarrem:2012wp} and depends on the definition of distance employed \cite{Albani:2006jq}.

\subsubsection{Shear:~~~}
Since the rationale behind this project is to test the Copernican principle we need to reconstruct a quantity that differs significantly in homogeneous and inhomogeneous models. For this we have chosen to evaluate the shear since it vanishes in FLRW universes and depends on the density only indirectly through its effect on the evolution of the other variables. Computing the magnitude of the shear $\sigma^2 = \frac{1}{2}\sigma_{ab}\sigma^{ab}$ with the metric \eqref{CIVPMet} and using the fact that $u_1 = u^0$ gives the following expression
\begin{eqnarray}
\sigma^2 = \frac{1}{12(u_1D)^4}\big[u_1^{\ 3}\left(2D\dot{D} - DW' + (2W+D)D' \right) \nonumber \\
+u_1^{\ 2}\left(-(DW + D^2)(u_{1})' - 2D^2 \dot{u}_{1} \right) - u_1DD' - D^2(u_{1})' \big]^2.
\end{eqnarray}
It is important to realise that the numerical errors resulting from the integration scheme will never yield a value of $\sigma^2 = 0$, even for homogeneous models. The best we can do is to verify, using a three level convergence test, that the shear associated with homogeneous models converges to machine tolerance as the grid is refined.\\
More problematically, we found that integrating the smoothed simulated FLRW data does not result in a $\sigma^2$ that converges to machine tolerance. This is most likely a result of using imperfect data. The limited number of available data points combined with the fact that uncertainty in realistic cosmological data is not simply Gaussian noise (see section~\ref{sec:SimDat}) means that the Gaussian process fails to correctly identify the underlying model. For this reason, and because it is practically impossible to prove that something is exactly zero, it is impossible to test the CP purely by inspecting the shear associated with the available data sets. For the purposes of this paper we propose a test in section~\ref{sec:test} that, although not completely model independent, allows us to discriminate between homogeneous and inhomogeneous models.

\subsection{Reconstructing a function and its derivatives} \label{sec:GaPP}
Given the raw data sets $D(z)$ and $H_{||}(z)$ we smooth the data using Gaussian process regression. A Gaussian process is a collection of random variables, any finite collection of which have a joint Gaussian distribution \cite{rasmussen2006gaussian}. The process is completely characterised by specifying its mean $m(x)$ and covariance function $k(x,\tilde{x})$. The mean and covariance function of a real process $f(x)$ are defined by
\begin{eqnarray}
m(x) &=& \mathbb{E}[f(x)], \\
k(x,\tilde{x}) &=& \mathbb{E}[(f(x) - m(x))(f(\tilde{x}) - m(\tilde{x}))],
\end{eqnarray}
where $\mathbb{E}[x]$ denotes the expectation value of $x$ with respect to a Gaussian distribution. An important result regarding Gaussian processes is that their derivatives are themselves Gaussian processes. It is this fact that enables us to reconstruct the derivatives of functions on the PNC0. A detailed explanation of how this can be achieved is given in \cite{Seikel:2012uu} and \cite{rasmussen2006gaussian}.\\
Although it is possible to specify non-zero mean functions for the process, we do not do so here as we do not want to favour any particular parametrisation over any other. Our prior belief about the data is that all parametrisations are equally likely\footnote{In all cases the posterior is not flat so that this prior is truly non-informative}. Phenomenologically this corresponds to imposing no physical constraints on the data and could lead to non-physical behaviour. The smoothing of the data is therefore followed up by a screening process (described in the next section) to extract the physically admissible combinations of data points. \\
The absence of a parametrization means that at the edges of the data set derivatives are poorly constrained. This gives rise to what we will refer to as edge effects. Special precautions are needed close to the origin to deal with artificial problems that arise as a result of edge effects.\\
For the covariance functions we have used the Mattern class of functions which take the form
\begin{equation}
k(x,\tilde{x}) = \frac{2^{1-\nu}}{\Gamma(\nu)}\left(\frac{\sqrt{2\nu}|x - \tilde{x}|}{l}\right)^\nu K_\nu \left(\frac{\sqrt{2\nu}|x-\tilde{x}|}{l}\right),
\label{CovFunc}
\end{equation}
where $\nu$ and $l$ are positive parameters and $K_\nu$ is a modified Bessel function. The Mattern class of functions are stationary and isotropic as they only depend on the magnitude $|x-\tilde{x}|$. Any process that uses this class of functions will be $k$-times mean square differentiable if and only if $\nu > k$. GaPP implements this class of functions with $\nu$ a half integer. We made use of \eqref{CovFunc} with $\nu = 5/2$\footnote{This function was chosen because it seemed to out perform others when reconstructing simulated data}, i.e.
\begin{equation}
 k(x,\tilde{x}) = \sigma_f^2 \exp\left[-\frac{\sqrt{5}|x-\tilde{x}|}{l}\right]\left(1 + \frac{\sqrt{5}|x-\tilde{x}|}{l}  + \frac{5|x-\tilde{x}|^2}{3l^2}\right).
\end{equation}
The hyper-parameters $l$ and $\sigma_f$ are known as the characteristic length scale and signal variance respectively. To get a truly Bayesian reconstruction of the function they need to be marginalised over. The relevant domain (i.e. the domain in which the likelihood function is significant) of $l$ and $\sigma_f$ is identified by finding the values, say $l^*$ and $\sigma_f^*$, that maximise the likelihood function. They are then initialised by specifying one hundred values on the interior of a regular grid defined by a box $2l^*$ by $2\sigma_f^*$. GaPP's \emph{mcmcdgp} module employs an affine invariant MCMC ensemble sampler (\emph{emcee} \cite{ForemanMackey:2012ig}) to explore the likelihood as a function of $(\sigma_f,l)$. It is then able to return correlated samples of the function and its first and second derivatives. These samples define the sample space $\mathcal{S}$ that is screened as described below.

\subsection{Screening process} \label{sec:screening}
The absence of a physical prior in the smoothing procedure gives rise to non-physical behaviours, especially at the edges of the data set. The null energy condition $\rho > 0$ provides a simple way to identify non-physical combinations of data from the sample space. Consider equation \eqref{density}. Clearly if $D'' > 0$ the null energy condition is violated. Combining the $\frac{dz}{d\upsilon}$ relation with successive applications of the chain rule makes it possible to write $D''$ on each PNC as
\begin{equation}
\frac{d^2 D}{d\upsilon^2} = 2(1+z)^3H^2\frac{dD}{dz} + (1+z)^4H\frac{dH}{dz}\frac{dD}{dz} + (1+z)^4H^2\frac{d^2D}{dz^2}.
\label{rnunu}
\end{equation}
Since $D$, $H$ and both of their first and second derivatives (as well as the covariances between them) are known in $\mathcal{S}$ it is possible to calculate the value $D''$ as a function of redshift. By repeatedly sampling from $\mathcal{S}$ we construct estimates for the distributions of $D, H$ and $\rho$ at each value of $z$ where we wish to reconstruct the function. A caveat of this approach is that the resulting distributions are not in general Gaussian. The mean and confidence intervals are therefore determined by calculating the area under the curve of each of them. In summary the screening process consists of the following steps:
\begin{enumerate}
\item For each redshift draw random samples from $\mathcal{S}$ using the output from GaPP.
\item Use the expression \eqref{rnunu} to test if the sample satisfies $D'' \leq 0$ and save the sample values if they do, otherwise discard them.
\item Repeat this process enough times to get a satisfactory approximation to the posterior distributions (the number of iterations required to get smooth functions will depend on the number of random variables in $\mathcal{S}$, in this case about 500 000 successful draws is sufficient).
\item Compute the area under the approximate distributions to extract the mean and relevant confidence intervals.
\item Move on to the next redshift value and repeat.
\end{enumerate}
Enforcing the null energy condition effectively places a prior on both data sets simultaneously. This raises an interesting point. The fact that LTB models only have two functional degrees of freedom means that it is not possible to specify a third arbitrarily. Any additional information has to be treated as a prior on what is already known. Density data should therefore not be treated as independent, instead it should be used as a prior on the other data sets. \\
The process as outlined above implements a lower bound on the density but places no restriction on the maximum allowed values. The reconstructed density is therefore likely to be over estimated and this is especially apparent close to the edges of the data sets. A considerable improvement should result if a model independent upper bound can be placed on the allowable density values.

\subsection{Considerations at the origin} \label{sec:origin}
Since $D \rightarrow 0$ as $\upsilon \rightarrow 0$, the presence of the $D^{-1}$ term in \eqref{density} poses numerical difficulties close to the origin. Due to a combination of edge effects and the lack of an upper bound on the density, its value tends to be grossly over estimated there. However, note that in spherical symmetry regularity at the vertex requires the density to behave like
\begin{equation}
\rho(\upsilon) = -2\frac{D'''(0)}{\kappa D'(0)} + O(\upsilon^2).
\label{originden}
\end{equation}
It is therefore possible to extrapolate its value at the origin by fitting a function $\rho(\upsilon) = \rho_0 + \rho_1 \upsilon^2$ to the data at a point where we expect edge effects to be negligible. To determine a decent fitting point we slide a fixed interval width of $\Delta \upsilon = 0.3$ along the function until the fit yields a value of $\rho' \geq 0$. The algorithm is therefore limited because it imposes a locally increasing density down the PNC. This is far from ideal but, to our knowledge, unavoidable with currently available data. The test of the CP presented in section~\ref{sec:test} does not use data in this region.\\
To avoid artificial instabilities a blending function of the form
\begin{equation}
\mbox{Blend}(\upsilon) = \frac{1}{2} + \frac{1}{2}\sin\left(\pi\left[\frac{\upsilon - \upsilon_{i}}{\upsilon_{e} - \upsilon_{i}}\right] - \frac{\pi}{2}\right),
\end{equation}
is used to smoothly blend the two solutions together. Here $\upsilon_{i}$ and $\upsilon_{e}$ are the values of $\upsilon$ at the start and end of the blending region respectively.

\section{Testing the Copernican Principle} \label{sec:Res}
Since we are limited to observing the universe from effectively a single space-time point, it is difficult to disentangle the degeneracy between temporal evolution and spatial variation. This is particularly evident in attempts to interpret the apparent late time acceleration of the universe as being due to either spatial inhomogeneity or dark energy \cite{Valkenburg:2011ty,Marra:2012pj}. Recent attempts to detect radial inhomogeneity in the class of LTB models including dark energy have revealed no statistically significant departures from FLRW geometry \cite{Valkenburg:2012td}. Instead of attempting a complete reconstruction of the metric within our causal past, the approach followed in \cite{Valkenburg:2012td} is to systematically falsify non-Copernican models and therefore rule out radial inhomogeneity as an alternative to dark energy. This approach is complementary to ours. 


\subsection{Simulating ``realistic" data}\label{sec:SimDat}
\begin{figure}
\includegraphics[width=\textwidth]{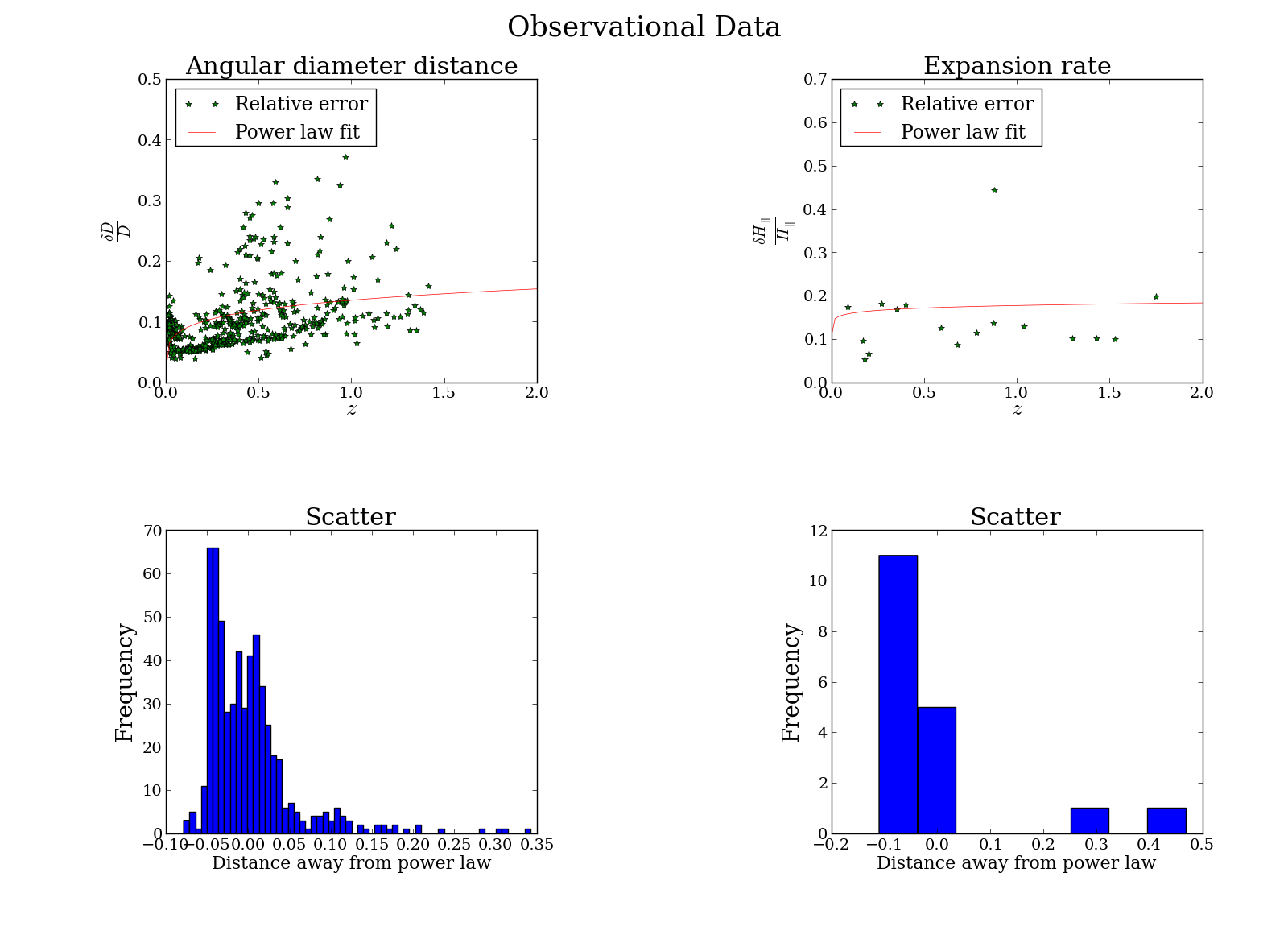}
\caption{The power law fit to the relative error of the observational data. Also shown is how the data are scattered about the power law.}
\label{ObsRelErr}
\end{figure}

\begin{figure}
\includegraphics[width=\textwidth]{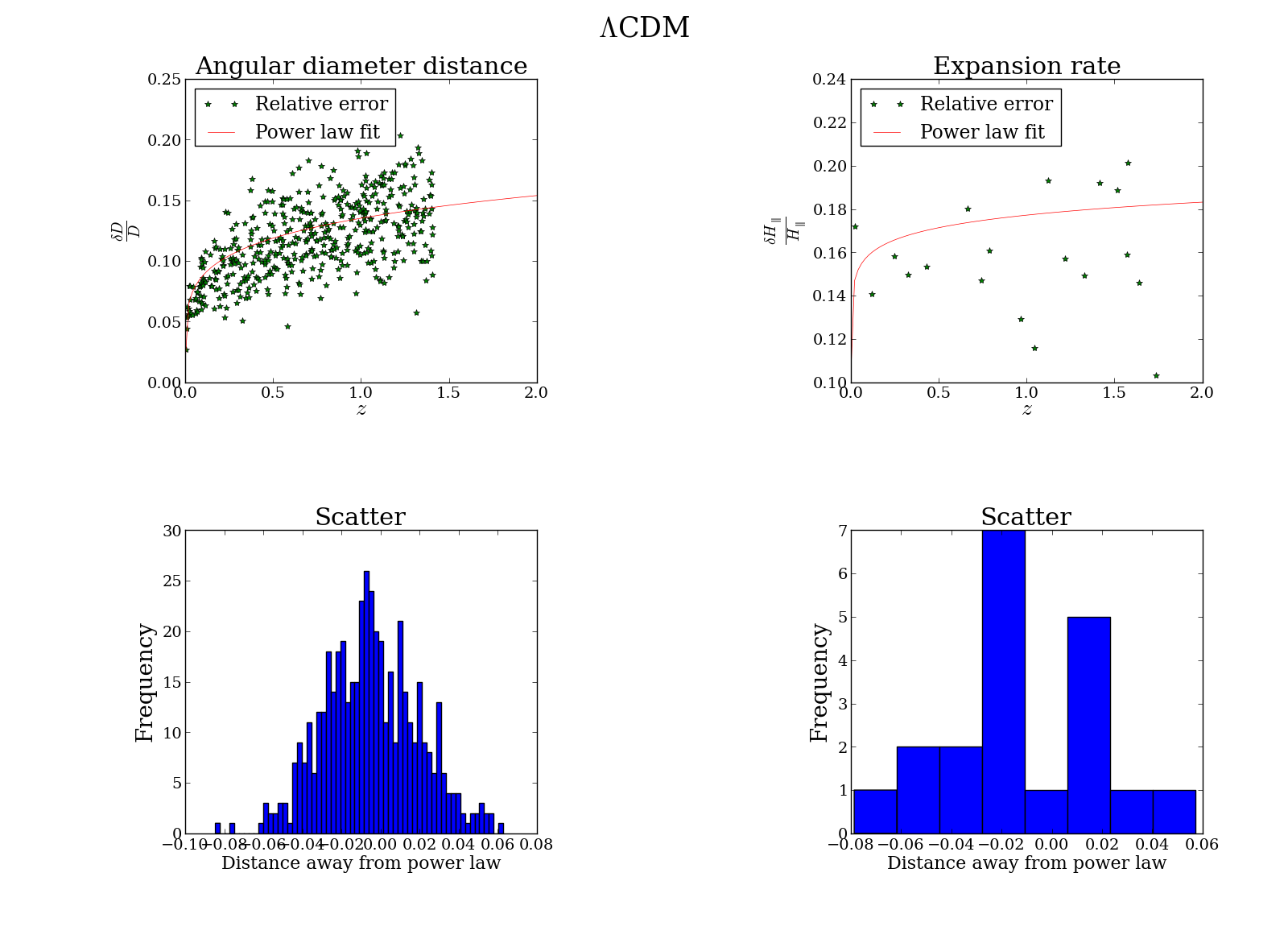}
\caption{This figure shows how the simulated $\LCDM$ data are scattered about the power law fit to the relative error of the observational data. Note that the observational data is more widely scattered and that the density of data points is greater close to the origin. This should be taken into account in realistic tests of the CP.}
\label{FLRWRelErr}
\end{figure}

\begin{figure}
\includegraphics[width=\textwidth]{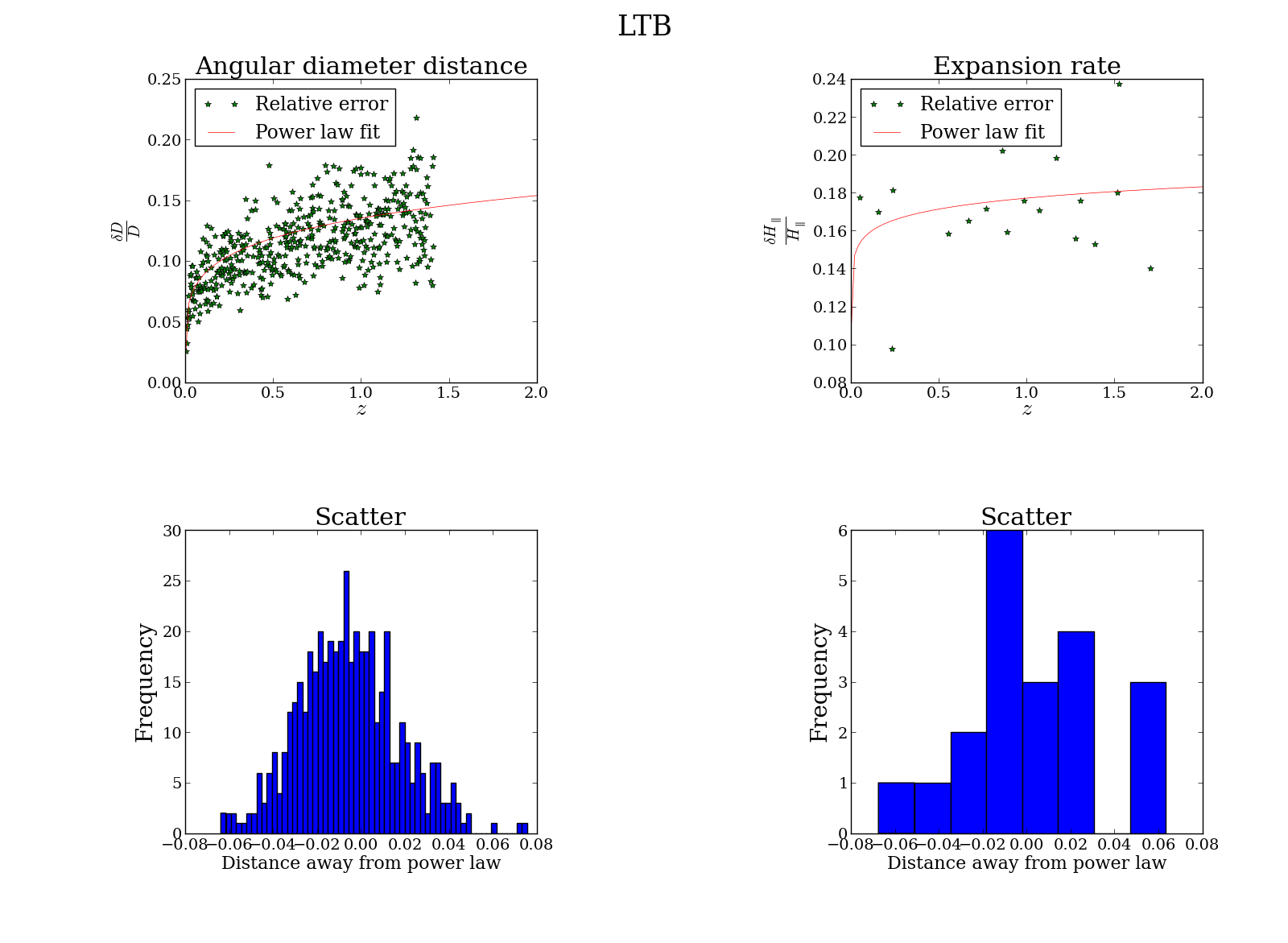}
\caption{This figure shows how the simulated LTB data are scattered about the power law fit to the relative error of the observational data. Ditto the remark in the caption of figure~\ref{FLRWRelErr} concerning observational data.}
\label{LTBRelErr}
\end{figure}

To ``calibrate" the algorithm it is necessary to simulate and compare $\LCDM$ and LTB data. There are a number of subtleties regarding this process that are worth discussing. In particular it is worth noting that realistic data sets are often significantly skewed and have non-monotonic cumulative probability distributions. It is known for example that low redshift observations are dominated by peculiar velocities that need to be measured and corrected for. It is also typical of flux limited surveys to observe that error grows with redshift. In this paper the aim is not to simulate realistic mock data for any particular survey and we assume that a power law in redshift $az^b$ is sufficient to model the average uncertainty/redshift trend. Further, because of experimental error, even if all factors are taken into account it is still unlikely that the data would consist of pure Gaussian noise. To investigate the degree to which the algorithm can cope with imperfect data the following simulating procedure is used:
\begin{enumerate}
\item Draw $n$ redshift values in the relevant range from a uniform distribution ($n$ is the number of data points simulated).
\item Suppose $F$ is the quantity being simulated and that $\bar{F}$ is its mean. Draw $N$ samples of $F$ using
\begin{equation}
F(z_n)=\bar{F}(z_n)\left(1 + az_n^{\ b} \mathcal{N}(0,I_N)\right).
\end{equation}
Here we have fitted the power law to $\frac{\delta F}{F}$ and $\mathcal{N}(0,I_N)$ is an uncorrelated Gaussian noise i.e. $I_N$ is an $N \times N$ identity matrix.
\item Sort the samples at each redshift and call the value at $N/2$ the mean, the value at $0.16 N$ the lower $1-\sigma$ value and use this to set the error bars on the data. \end{enumerate}
Simulating data in this way makes it possible to test a number of factors (see figures \ref{ObsRelErr}-\ref{LTBRelErr}). Firstly it should be noted that the value of $N$ controls the degree of Gaussianity in the data, as $N$ is finite the distribution of error is not perfectly Gaussian. In this paper $N = 51$. \\
Again, let us stress that the data sets thus simulated are not realistic: it is simply a way to impose imperfections on the data to test the algorithm presented in this paper. For more realistic tests of the CP it will be necessary to simulate data in a more sophisticated way.\\
Both $D(z)$ and $H_\|(z)$ data have been simulated using this procedure. The maximum redshift for each simulation is set by the real data and we use the same number of data points i.e. 580 $D(z)$ points with $z \in \{0,1.414\}$ and 18 $H_\|(z)$ points with $z \in  \{0,1.75\}$.\\

\subsection{LTB}
Using the form of the metric \eqref{LTBmetric2} the analogue of the Friedmann equation in this space-time is given by
\begin{equation}
H_{\perp}^2 = \frac{M}{R^3}-\frac{k}{R^2},
\label{H}
\end{equation}
where $M=M(r)$ is an integration function playing the role of effective mass, and the locally measured energy density is
\begin{equation}
8 \pi G \rho(t,r) = \frac{\partial_r(M r^3)}{R^2\partial_r R}.
\end{equation}
Introducing dimensionless density parameters for the CDM and curvature by analogy with the FLRW models i.e.
\begin{equation}
 \omegko(r)=-\frac{k(r)}{\Hperpo^2(r)}, \quad \mbox{and} \quad \omegmo(r)=\frac{M(r)}{\Hperpo^2(r) },
\end{equation}
the Friedmann equation takes on a more familiar form:
\begin{equation}
\frac{\Hperp^2(t,r)}{\Hperpo^2(r)}=\omegmo(r)\left(\frac{R_0(r)}{R(t,r)}\right)^3 + \omegko(r)\left(\frac{R_0(r)}{R(t,r)}\right)^2, \quad \omegmo(r)+ \omegko(r)=1.
\end{equation}
Integrating the Friedmann equation from the time of the big bang $t_B = t_B(r)$ to some later time $t$ yields the age of the universe at a given $(t,r)$:
\begin{equation}
\label{solA}
t - t_B(r) = \frac{1}{\Hperpo(r)}\int^{\frac{R}{R_0}}_{0}\frac{d x}{\sqrt{\omegmo (r)x^{-1} + \omegko (r)}}\,.
\end{equation}
At this stage there are two free functional degrees of freedom viz. $\omegmo(r)$ and $t_B(r)$. Ignoring decaying modes (\cite{silk}, \cite{zibin}) we set $t_B = 0$ so that the model evolves from FLRW. As a result the age of the universe is constant and equal to the time today $t_0$. So, by solving (\ref{solA}) for $\Hperpo (r)$, we have that:
\ba
\Hperpo(r)=\left\{
\begin{array}{cr}
 \displaystyle \frac{- \sqrt{-\omegko }+\omegmo\sin^{-1}\sqrt{-\frac{\omegko}{\omegmo}} } {t_0\left(-\omegko\right)^{3/2}}\,   &  \omegko<0 \\[4mm]
\displaystyle  \frac{2}{3t_0}   & \omegko=0   \\[2mm]
\displaystyle  \frac{\sqrt{\omegko } - \omegmo\sinh^{-1}\sqrt{\frac{\omegko}{\omegmo}} } {t_0\omegko^{3/2}}   &  \omegko>0
\end{array} \;.
\right.
\ea
The standard way to solve for $R(t,r)$ is to introduce an additional parameter $\eta$ in which case the solution is
\begin{eqnarray}
R(t,r) &=& \frac{\omegmo}{2\omegko} \left(\cosh\eta - 1\right)R_0(r), \\
\Hperpo t &=& \frac{\omegmo}{2\omegko^{3/2}}\left(\sinh \eta - \eta \right),
\end{eqnarray}
where we will work in the gauge $R_0(r) = r$. The model is therefore completely determined by specifying a density profile. In what follows the matter density is chosen to obey
\begin{equation}
\omegmo (r) = \omegmin + (\omegmout - \omegmin )\left(\frac{1-\tanh[(r-r_0)/2\Delta r]}{1+\tanh[r_0/2\Delta r]}\right),
\end{equation}
where $\omegmin$ and $\omegmout$ are the values of $\omegmo$ at the centre of the void and at infinity, respectively. The parameter $r_0$ characterizes the size of the void while $\Delta r$ characterises the transition to uniformity. The asymptotic behaviour is specified by fixing $\omegmout=1$ so that the spacetime is asymptotically flat. For the other remaining free parameters we use the best fit values obtained in \cite{GarciaBellido2008nz} (i.e. $\omegmin = 0.13$ and $\Hperpo(0) = 64 kms^{-1}Mpc^{-1}$) to which the reader is also referred to for further details regarding this model (however see also \cite{Alnes:2005rw}).\\
Now that the model has been specified observables are determined as a function of redshift by solving
\begin{equation}
\frac{dt}{dz} = -\frac{1}{(1+z)\Hpar} \quad \mbox{and} \quad \frac{dr}{dz} = \frac{\sqrt{1-\kappa r^2}}{(1+z)\Hpar\partial_r R},
\label{tzrz}
\end{equation}
where $H_\|(t,r)=H_\|(t(z),r(z))$ and the area distance is given by $D(z)=R(t(z),r(z))$.
Finally, the metric \eqref{LTBmetric2} gives the shear in this model as
\begin{equation}
\sigma^2 = \frac{\left(X \partial_t R - R\partial_t X \right)^2}{3 X^2 R^2}.
\label{LTBshear}
\end{equation}

\begin{figure}
\includegraphics[width=1\textwidth]{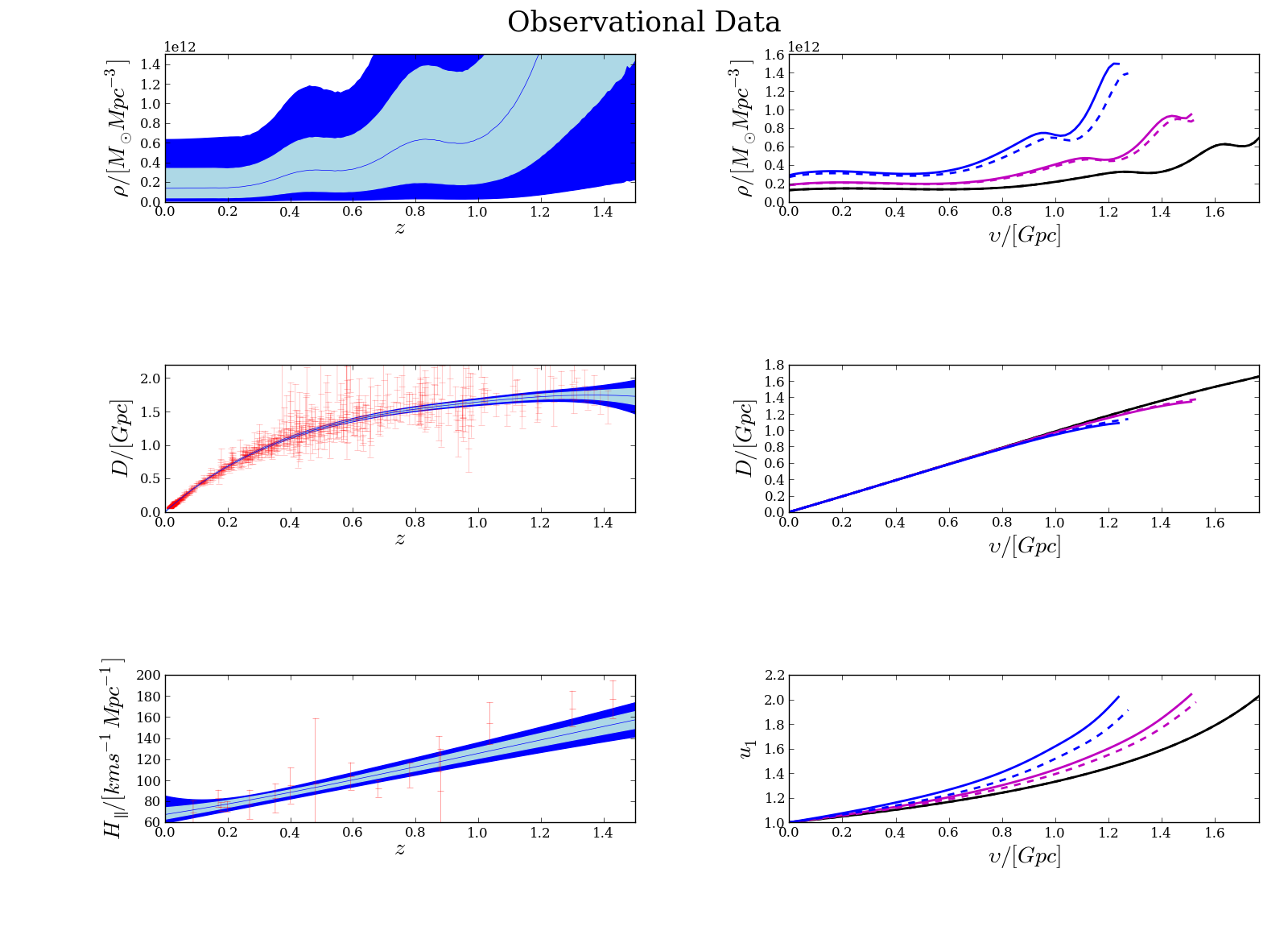}
\caption{Left) The $1/2-\sigma$ reconstruction of the observational data on the PNC0 as a function of $z$. Blue lines correspond to the most probable values that are passed into to the integration scheme. Note that for $z < 0.2$ the density has been extrapolated as explained in section~\ref{sec:origin}. Right) Here we show the result of integrating the most probable function values on the left with $\Omega_\Lambda$ set to both $0.7$ and $0$. The solid lines correspond to $\Omega_\Lambda = 0$ and the dashed ones to $\Omega_\Lambda = 0.7$. We show the results on three PNCs. The black lines corresponds to the PNC0 at $w = 13.6$, the magenta lines to the PNC defined by $w = 11.9$ and the blue lines to that of a PNC defined by $w = 10.2$. We could go back further in time by refining the grid. }
\label{fig:ObsDat}
\end{figure}

\begin{figure}
\includegraphics[width=1\textwidth]{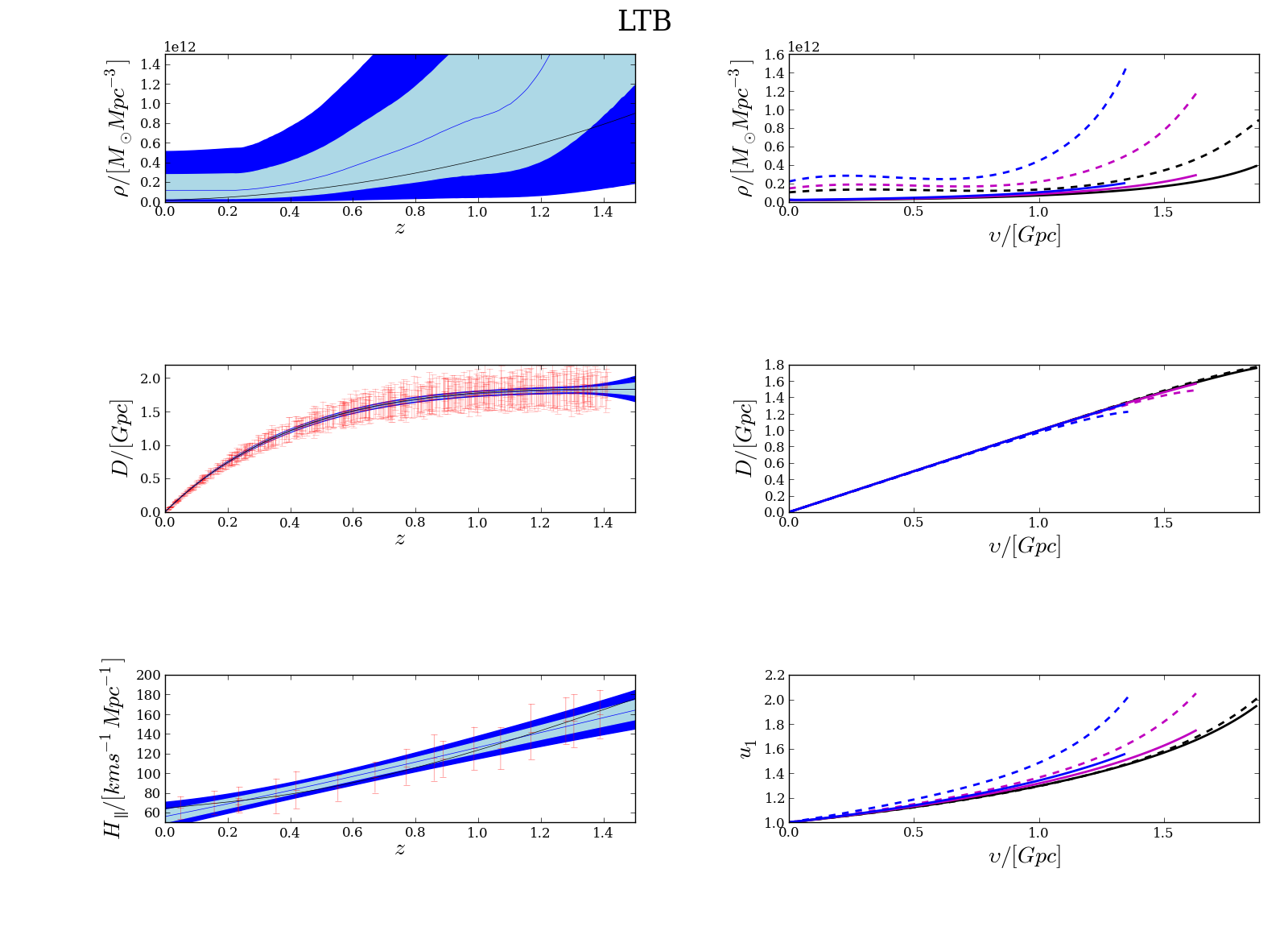}
\caption{Left) The reconstructed LTB data on the PNC0 as a function of $z$. Black lines correspond to the underlying LTB model, blue lines again correspond to the most probable values and we show the $1/2-\sigma$ contours. The density below $ z \approx 0.2$ is again extrapolated. Note that, as discussed in section~\ref{sec:screening}, the reconstructed density is grossly over-estimated. These figures illustrate the importance of obtaining independent density data, especially at low redshifts. Right) The result of integrating the most probable function values on the left. The solid lines correspond to the model and dashed lines to the most probable reconstructed function values. Black lines again correspond to the PNC0 at $w = 13.6$, the magenta lines to the PNC defined by $w = 11.9$ and the blue lines to that of a PNC defined by $w = 10.2$.}
\label{fig:LTBDat}
\end{figure}

\begin{figure}
\includegraphics[width=1\textwidth]{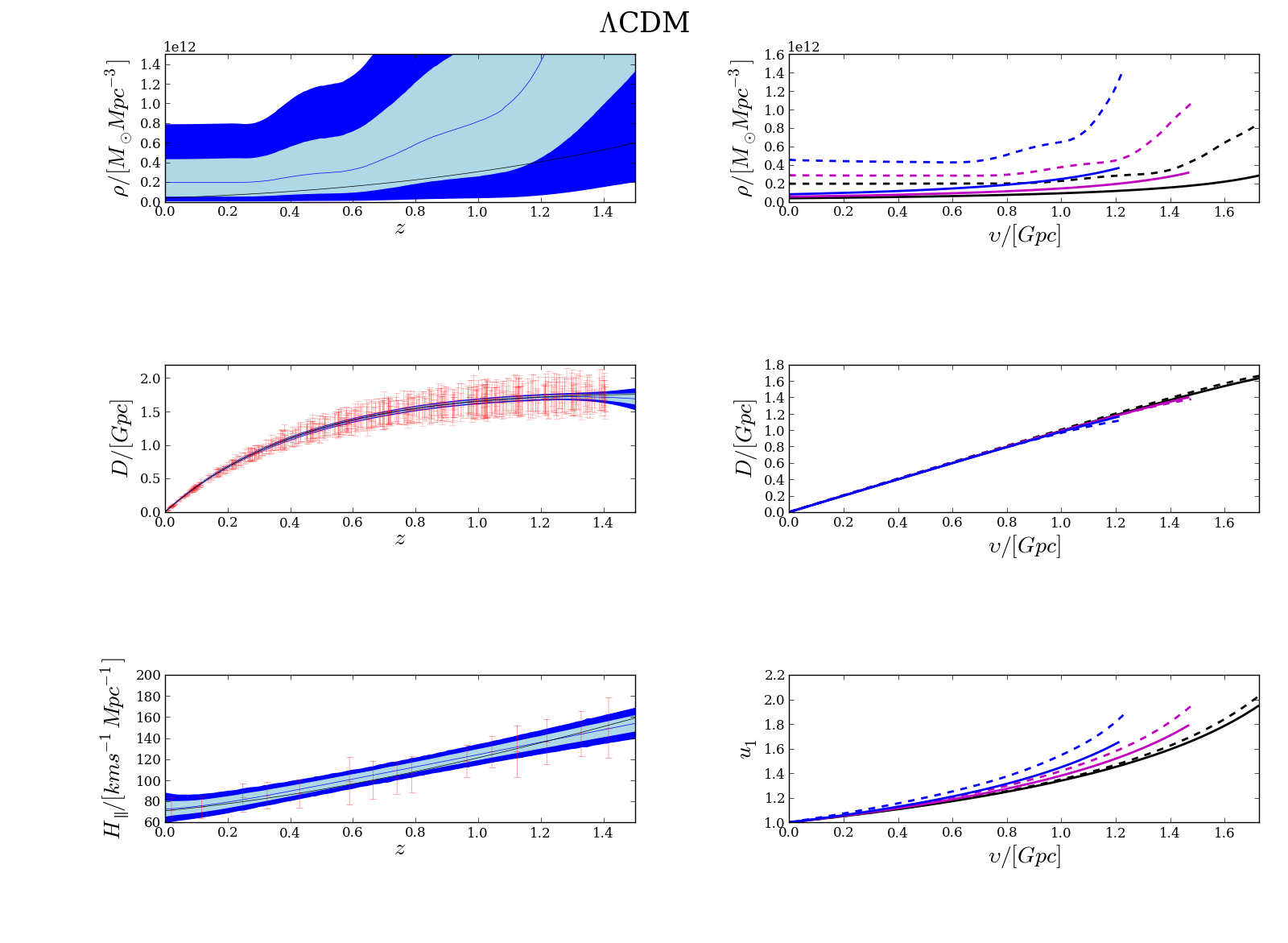}
\caption{This figure is identical to figure~\ref{fig:LTBDat} but for simulated $\LCDM$ instead of LTB.}
\label{fig:FLRWDat}
\end{figure}

\subsection{FLRW}
Using the FLRW form of the metric
\begin{equation}
ds^2 = -dt^2 + a(t)\left(\frac{dr^2}{\sqrt{1-k r^2}} + r^2 d\Omega^2\right),
\end{equation}
we also simulate data for the a $\LCDM$ model specified by
\begin{equation}
\Omega_{m0} = 0.3, \quad \Omega_{K0} = 0, \quad \Omega_{\Lambda 0} = 0.7, \quad H_0 = 71 \frac{km}{sMpc}.
\label{LCDMPar}
\end{equation}
With this choice of $\Omega_K = 0$ the angular diameter distance is simply
\begin{equation}
D(z) = \frac{1}{(1+z)}\int_0^z \frac{dz}{H}.
\end{equation}
Now the Hubble rate obeys the dimensionless form of the Friedmann equation
\begin{equation}
H(z) = H_0\left(\Omega_m(1+z)^3 + \Omega_\Lambda \right),
\end{equation}
which gives $\rho(z) = 3 H(z)^2 \Omega_{m0}/8\pi$ and $\Lambda = 3 H^2_0 \Omega_{\Lambda 0}$. Normalising the current day scale factor to $a(t_0) = 1$ the current age of the universe is
\begin{equation}
t_0 = \frac{2\sinh^{-1}(1/\Omega_m - 1)}{3H_0\sqrt{\Omega_\Lambda}}
\end{equation}

\begin{figure}
\centering
\includegraphics[width=1.1\textwidth]{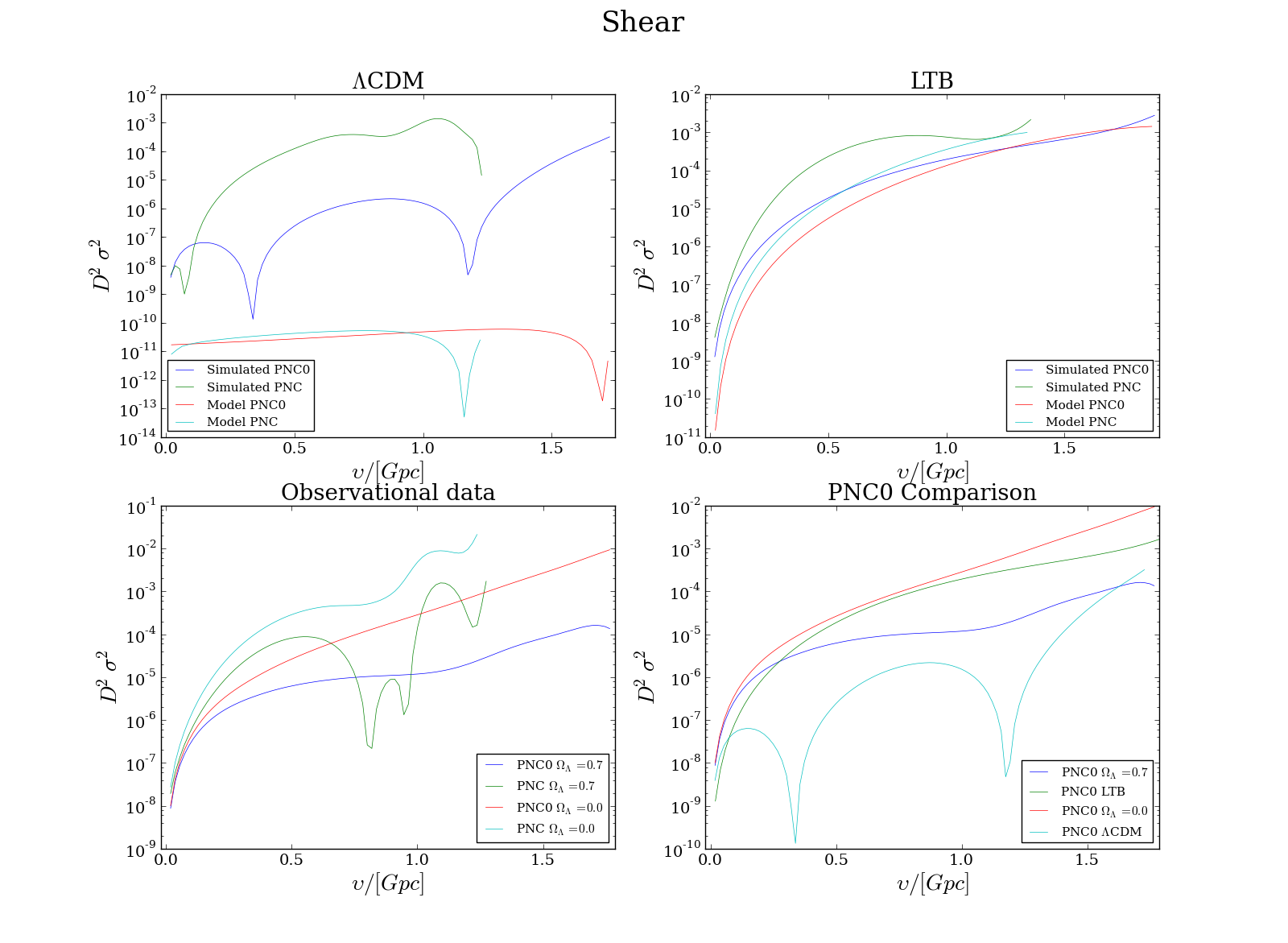}
\caption{This figure illustrates $D^2\sigma^2$ resulting from integrating the most probable reconstructed functions for the various simulations (note the log scale on the y axis) on two different PNCs. Top left: How the shear for simulated $\Lambda$CDM data compares to that of the model. Notice that the shear from the reconstructed simulated data is far greater than that of the model (which is purely due to numerical error). This is the intrinsic or ``artificial" shear that we should expect from imperfect data. Top right: The same comparison for LTB data. Since the shear in the LTB model is significantly greater than the numerical error of the integration scheme the numerical errors are insignificant, hence the improvement in the reconstruction. Bottom left: The shear for observational data with two different values for $\Lambda$ but everything else identical. Clearly the value of $\Lambda$ plays a significant role in the evolution of the shear. Bottom right: A comparison of the observational data with the two models. Notice how well the reconstructed shear of the LTB data compares to the observational shear computed with $\Lambda = 0$. Also notice that the shear from observational data with $\Omega_\Lambda = 0.7$ seems to be orders of magnitude larger than the reconstructed $\LCDM$ data.}
\label{fig:Shear}
\end{figure}

\subsection{The test}\label{sec:test}
There are fundamental differences between LTB and FLRW models that can be used to distinguish them. It is known for example that, below $z \approx 2$, redshift drift has the opposite sign in LTB and FLRW models \cite{Yoo:2010hi,Yoo:2008su}. However, this is not necessarily true if the cosmological constant is allowed to be non-zero. The test that we present in this paper therefore uses the (matter) shear instead. To illustrate how the shear can be used to discriminate between homogeneous and inhomogeneous models the following measure is introduced
\begin{equation}
\mathcal{P}_1 = \frac{\int_{\upsilon_{min}}^{\upsilon^{max}} |\sigma^2 - \bar{\sigma}^2_{FLRW}| d\upsilon}{\int_{\upsilon_{min}}^{\upsilon^{max}} |\sigma^2_{test} - \bar{\sigma}^2_{FLRW}|d\upsilon}.
\label{measure}
\end{equation}
Here $\upsilon_{min}$ and $\upsilon_{max}$ are the limits to where we can trust the reconstructed data and a bar refers to the shear of the underlying model computed using the integration scheme. The quantity $\sigma^2$ is the reconstructed shear of the data set being tested. Then $\sigma^2_{test}$ is the shear associated with the simulated data of the (inhomogeneous) model that the data is being tested against. If $\mathcal{P}_1 \gtrsim 1$ the shear associated with the observational data is of the same order of magnitude, or larger than, that of the test model. As $\mathcal{P}_1 \rightarrow 0$ the evidence that $\sigma^2$ corresponds to a homogeneous model increases. However note that the value of $\mathcal{P}_1$ is meaningless as it stands, since $\sigma^2$ from realistic data will inevitably be quite different from zero. To ``calibrate'' \eqref{measure} for the test model we also define a second measure as
\begin{equation}
\mathcal{P}_2 = \frac{\int_{\upsilon_{min}}^{\upsilon^{max}}|\sigma^2_{test} - \sigma^2_{FLRW}|^2 d\upsilon}{\int_{\upsilon_{min}}^{\upsilon^{max}} \sigma^2_{test}d\upsilon}.
\label{calibrator}
\end{equation}
If the data is sufficient to discriminate between the models then $\mathcal{P}_2 \rightarrow 1$, whereas a value closer to zero tends to point to the contrary. Simulating data for both the test and FLRW models and evaluating $\mathcal{P}_2$ then gives an indication of what data would be sufficient (at least with this methodology) to discriminate between them. Further, for the value of $\mathcal{P}_2$ to have statistical significance, it is important to test the value of $\mathcal{P}_2$ for many different simulated data sets. Then, if the data truly are sufficient to discriminate between the models, the value of $\mathcal{P}_2$ should not differ significantly between simulations.\\
Notice that if the data favours the inhomogeneous test model we would expect a value close to one for both $\mathcal{P}_1$ and $\mathcal{P}_2$. A value for $\mathcal{P}_1$ close to zero could indicate that the data favours a homogeneous model if $\mathcal{P}_2$ is close to one. \\
Note that the result of this test is only a comparison of expected magnitudes. We do not claim that it is able to identify the underlying cosmological model. The models are introduced to provide a standard of comparison and the observational data should be confronted with as many of them as possible. Given the limitations of cosmological data it might be impossible, at least within this framework, to confirm that the shear is zero (this is akin to proving that the spatial curvature is identically zero).

\subsection{Results}
Figures~\ref{fig:ObsDat}-\ref{fig:FLRWDat} show the reconstructed function on the PNC0 as well as the result of integrating the most probable function reconstructions. Due to edge effects we truncate the functions beyond $z = 1.1$ and only perform the integrations for for $z \leq 1.1$. Note the effect of $\Lambda$ in the integrations shown on the right of figure~\ref{fig:ObsDat}. Since everything else is identical, it is clear that the evolution history is significantly different for different values of $\Omega_\Lambda$. The seeds of inhomogeneity might therefore be more evident on previous PNCs. Unfortunately, as illustrated by the value of $\mathcal{P}_2$ in the interior of the PNC below, currently available data does not permit statistically robust discriminations between homogeneous and inhomogeneous models in the interior.\\
In figure~\ref{fig:Shear} we compare the quantity $D^2\sigma^2$ for the various simulations (where we use $D^2\sigma^2$ instead of $\sigma^2$ because it is dimensionless). Numerical error grows both with increasing $\upsilon$ and as the integration proceeds towards the interior of the PNC \footnote{We have verified that the source of this error is purely numerical by doing a three level convergence test and showing that it converges to zero.}. In particular figure~\ref{fig:Shear} (bottom left) shows the shear that results from integrating the real data with both $\Omega_\Lambda = 0$ and $\Omega_\Lambda = 0.7$. Since the density below $z \approx 0.2$ has been extrapolated, we only perform the test in the range $0.3 \leq z < 1.1$ (equivalently $\upsilon_{min} = 0.8$ and $\upsilon_{max} = 1.7$). Performing the test described above on the PNC0 we find that:
\begin{itemize}
	\item For $\Omega_\Lambda = 0.7: \quad \mathcal{P}_1 = 0.06, \quad \mathcal{P}_2 = 0.99$,
	\item For $\Omega_\Lambda = 0.0: \quad \mathcal{P}_1 = 1.55, \quad \mathcal{P}_2 = 0.99$.
\end{itemize}
Performing the same test on the earliest PNC (at a value of $w = 10.2$) but with $\upsilon_{min} = 0.8$ and $\upsilon_{max} = 1.2$ we find that:
\begin{itemize}
	\item For $\Omega_\Lambda = 0.7: \quad \mathcal{P}_1 = 0.78, \quad \mathcal{P}_2 = 0.50$,
	\item For $\Omega_\Lambda = 0.0: \quad \mathcal{P}_1 = 6.32, \quad \mathcal{P}_2 = 0.50$.
\end{itemize}
The significantly lower value of $\mathcal{P}_2$ on the earliest PNC might be due to the effect that the density has on the evolution of other quantities. This emphasises the need for independent density data, especially at low redshifts. Further, these results tend to indicate that the value of $\Lambda$ could be the deciding factor in discriminating between homogeneous and inhomogeneous models.

\section{Discussion and Conclusion}\label{sec:discussion}
We have presented an algorithm that is in principle capable of reconstructing the metric inside our PNC for any spherically symmetric dust dominated universe. Some advantages of the algorithm include its insensitivity to the apparent refocussing of the PNC and the Bayesian methodology employed to smooth observations on our PNC0. Yet, as it stands, the algorithm is currently incapable of providing robust statistical information about the interior of the PNC and, as illustrated, currently available data does not allow for a fully satisfactory test of the CP. \\
Probably the most difficult aspect of the observational cosmology programme is obtaining model independent data. This is especially true for the expansion rate data we have employed since different models of galaxy evolution yield different results. This programme will therefore benefit from future science missions such as LADUMA \cite{Holwerda:2011kd} that aim to further our understanding of galaxy evolution. Moreover we hope that this article elucidates the importance of performing the analysis in a model independent way i.e. without the Copernican assumption. \\
The algorithm employed to reconstruct the density is necessitated by the lack of model independent density data. As can be seen from figures~\ref{fig:LTBDat} and \ref{fig:FLRWDat} the reconstruction fails close to the edges and is always slightly over-estimated. This is likely due to the lack of an upper bound on the allowable values of $\rho$. It remains to be seen what improvement can be achieved by incorporating data from galaxy and galaxy cluster surveys. As pointed out in section~\ref{sec:screening}, any such data should be considered strictly as a prior on the other data sets.\\
Since the analysis presented here merely serves as an illustration of principle, we have only used the most probable reconstructed function values to ``test" the CP. These results therefore do not carry any statistical significance. However the algorithm has been developed with an extension in mind. The idea is based on drawing function realisations of $D(z)$ and $H_\|(z)$ instead of sampling redshift by redshift. The successful realisations (i.e. those with $\rho > 0$) could then be used to reconstruct the density (using numerical derivatives where needed). Performing the integration for each set of function realisations then provides an estimate of the posterior of the metric functions inside the PNC. This will also allow us to reconstruct the distributions of $\mathcal{P}_1$ and $\mathcal{P}_2$. Further, note that for each function realisation of $H_\|(z)$ the $\upsilon(z)$ relation is exact. The problem of having uncertainty in $\upsilon(z)$ can therefore be sidestepped altogether (values can be reconciled post-integration by interpolating). Such a procedure would also make it possible to marginalise over the value of $\Lambda$. Therefore, as promised, many of the difficulties mentioned in the paper can be overcome by drawing function realisations instead of sampling point by point. It remains to be seen how computationally viable such a procedure would be. Given the non-physical nature of Gaussian processes it might be that blind function realisations hardly ever pass the null energy condition and more sophisticated sampling techniques have to be employed. These issues are left for future research. \\
We have insisted from the outset that the data considered in this paper are not sufficient to test the CP. In our approach this is mainly due to the fact that we do not have a model independent value for $\Lambda$. This will change when redshift drift data becomes available. To illustrate, suppose that redshift drift data (i.e. $\dot{z}$) is available from observations and can be written in terms of $\upsilon$ through the $z(\upsilon)$ relation. The definition of $u^1$ then gives
\begin{equation}
u^1 = \frac{d\upsilon}{d\tau} = \frac{dw}{d\tau}\frac{d\upsilon}{dw} = (1+z)\frac{d\upsilon}{dz}\dot{z}.
\label{u1ForLam}
\end{equation}
Using $u_0 = -\left(1+\frac{W}{D}\right)u^0 + u^1$ with \eqref{u1ForLam}, and the fact that $u_1 = u^0$ gives
\begin{equation}
u_0 = \left(1+\frac{W}{D}\right)u_1 + (1+z)\frac{d\upsilon}{dz}\dot{z}.
\label{u0ForLam}
\end{equation}
Substituting \eqref{u0ForLam} into the normalisation condition \eqref{IntNorCon} gives $W(z)$ (and through the $z(\upsilon)$ relation $W(\upsilon)$), on the PNC as
\begin{equation}
W(z) = D\left(\frac{1}{3(u_1)^2} - \frac{2(1+z)}{3u_1}\frac{d\upsilon}{dz}\dot{z}\right).
\end{equation}
Since $D, W$ and $\dot{D} = \frac{dD}{dz}\dot{z}$ as well as, through numerical differentiation, their derivatives towards $\upsilon$ are known, the cosmological constant can now be obtained by rearranging \eqref{WNU}
\begin{equation}
\Lambda = \frac{1}{2D}\left(\frac{W}{D}D'' +4\dot{D}' + 2\kappa(u_0u_1\rho - \frac{1}{2}\rho) - W''\right).
\end{equation}
Since $\Lambda$ is expected to be constant this expression only needs to be evaluated at a single point, alternatively evaluating the expression at multiple points provides a test for the constancy of $\Lambda$.\\

\section{Acknowledgements}
The authors wish to thank Marina Seikel and Christopher Clarkson for many useful discussions. HLB also wishes to thank Denis Pollney for his assistance with some of the technical aspects he encountered during the course of his MSc. The financial assistance of the South African Square Kilometre Array project (SKA SA) towards HLB's research is hereby acknowledged. Opinions expressed and conclusions arrived at are those of the author and are not necessarily to be attributed to the SKA SA (www.ska.ac.za).  

\section{References}

\bibliographystyle{unsrt}
\bibliography{Whats_Inside_the_Cone}

\end{document}